\documentclass[a4paper,11pt]{article}
\usepackage{latexsym}	
\usepackage{amsmath}
\usepackage{amsthm}

\usepackage{ amssymb }
\def\ba{\begin{eqnarray}}
\def\ea{\end{eqnarray}}

\def\ba{\begin{eqnarray}}
\def\ea{\end{eqnarray}}

\def\lb{\label}
\def\be{\begin{equation}}
\def\ee{\end{equation}}

\theoremstyle{plain}

\begin{document}
\baselineskip0.25in
\title{The appearance of non trivial torsion for some Ricci dependent theories in the Palatini formalism}
 \author{ Juliana Osorio \thanks{Instituto de Matem\'atica Luis Santal\'o (IMAS), UBA CONICET, Buenos Aires, Argentina
juli.osorio@gmail.com.}   and Osvaldo P. Santill\'an\thanks{Instituto de Matem\'atica Luis Santal\'o (IMAS), UBA CONICET, Buenos Aires, Argentina
firenzecita@hotmail.com and osantil@dm.uba.ar.}}

\date {}
\maketitle

\begin{abstract}
As is known from studies of gravity models in the Palatini formalism, there exist two inequivalent definitions of the generalized Ricci tensor in terms of the generalized curvature namely, 
$\widetilde{R}_{\mu\nu}=R^\rho_{\mu\rho\nu}$ and $R^\mu_{\nu}=g^{\alpha\beta}R^\mu_{\alpha\nu\beta}$. A deep formal investigation of theories with lagrangians of the form $L=L(\widetilde{R}_{(\mu\nu)})$ was initiated in  \cite{olmo}. In that work, the  authors leave the connection free, and find out that the torsion only appears as a projective mode. This  agrees  with  the widely employed condition of  vanishing
torsion in these theories as a simple gauge choice. In the present work the complementary scenario is studied namely, the one described by a  lagrangian that depends on the other possible Ricci tensor $L=L(R_{(\mu\nu)})$. 
The torsion is completely characterized in terms of the metric and the connection, and a rather detailed description of the equations of motion is presented. It is shown that these theories are non trivial even for $1+1$ space time dimensions, and admit non zero torsion even in this apparently simple case. It is suggested that to impose zero torsion by force may result into an incompatible system. In other words, the presence of torsion may be beneficial for insuring that the equations of motion are well posed. The results of the present paper do not contradict the ones of \cite{olmo}, as the underlying theories are inequivalent.

 \end{abstract}

\section{Introduction}

The phenomenology of modified theories of gravity is of particular interest in the context of inflation, bouncing cosmology, and dark energy \cite{odintsov}. One of the goals
of modified gravity is to find a self-consistent theoretical description describing simultaneously the early-time and late-time acceleration of the universe. It seems plausible the description of very short or very large scale physics requires a modification of the Einstein-Hilbert lagrangian, thus modified gravity theories are of interest in this context. 

There is however, an interesting subtlety related to the possible generalizations of the Einstein equations. As is well known, the Einstein equations of motion arise 
from the Einstein-Hilbert action regardless the formalism employed is the metric or the Palatini one. In the first formalism, the variation of the action is achieved by considering
the metric components as the independent degrees of freedom. In the second, the metric and the connection are considered as independent variables. For a modified
gravity theory instead, the resulting equations of motion are inequivalent when these two formalism are applied. Thus, for a given guess of a modified scenario, there are at least two set possible phenomenological consequences to be understood.

There are several motivations for focusing in the Palatini formalism, in which the metric and the connection are taken as independent variables. An example is the Stelle model of gravity \cite{Stelle}, which has the remarkable feature of being renormalizable, but it contains Ostrogradskyi ghosts or instabilities in the metric formulation. Thus, a Palatini analysis for this model is of particular interest. There exist several modified gravity theories with a wide range of applications in the context of inflation and dark matter, as reviewed in \cite{odintsov} and references therein. Much of these models have been extensively studied in the metric formalism. However, there exist also works devoted to the Palatini formulations of these alternative gravity theories. For example, the Palatini formulation for minimal coupled fields can be found in \cite{BeltranJimenez} and  the analysis of Born-Infeld type of theories was presented in \cite{olmo}-\cite{Fiorini4}. 
In addition, literature studying the Palatini formalism in $f(R)$ and $f(R, R_{\mu\nu} R^{\mu\nu})$ theories of gravity  can be found in \cite{Capozziello}-\cite{Olmob}. In particular, an interesting result was reported in  \cite{Capozziello}, \cite{Sotiriou}, \cite{Olmolight} for the dynamics of $f(R)$ theories. It was shown in these references that the torsionless Palatini formulation of these theories
yield the same equations as the metric compatible ($\nabla^\Gamma_\mu g_{\alpha\beta}=0$) formulation. A preliminary work in that direction is \cite{volovic}.

The results just described seem to indicate a close relation between torsion and non-metricity in metric-affine theories of gravity, at least for some type of modified gravity scenarios. 
These studies were pushed forward in particular in \cite{olmo} for Einstein-Born-Infeld scenarios and other related gravity models, examples are found in  \cite{nonminimalscalar}-\cite{Jimenez}, \cite{BIE}-\cite{Bambi} and references therein. The lagrangian considered in \cite{olmo} is of the form $L=L(\widetilde{R}_{(\mu\nu)})$ with $\widetilde{R}_{\mu\nu}=R^\rho_{\mu\rho\nu}$ . One of the findings of this work is the presence of a projective symmetry
and that the torsion only
appears as a projective mode. This result justifies the procedure of setting vanishing
torsion in these theories as a simple gauge choice.

The results presented in \cite{olmo} are of interest from the theoretical and phenomenological point of view, as is a quite general statement about the mathematical structure of these modified gravity theories. However, the result is not fully general due to the following detail. The analysis these authors make is based on the choice $\widetilde{R}_{\mu\nu}=R^\rho_{\mu\rho\nu}$ for the Ricci tensor. The point is that, if the connection is not the standard Levi-Civita, then there exist another inequivalent choice for the curvature $R^\mu_{\nu}=g^{\alpha\beta}R^\mu_{\alpha\nu\beta}$. Both choices reduce to the standard curvature tensor for the Levi-Civita case, otherwise they are expected to be different objects. Thus, the most general lagrangians to be considered in order to make generic statements have the functional form $L=L(a R_{(\mu\nu)}+b \widetilde{R}_{(\mu\nu)})$  with $a+b=1$. This will clearly results in more complicated equations of motion.

The present work is intended to make a preliminary step in the generalization described above. The results of \cite{olmo} are related to the choice $a=0$, $b=1$. Instead, the present work considers the complementary case $a=1$ and $b=0$. This means that the lagrangians to be considered here are of the form $L=L(R_{(\mu\nu)})$. The corresponding equations are derived and simplified conveniently. The resulting theories seem quite different, in particular, it is shown that a non trivial torsion is turned on in this case. The resulting system seems to be more complex than in the case considered in \cite{olmo}. Some simple examples are worked out in 1+1 space time dimensions. It is found that even in these simple scenarios, a non trivial torsion may appear. In addition, it is argued that the torsion may be a fundamental object for the system to admit solutions. In authors opinion, to make steps toward the full generalization is of theoretical and phenomenological relevance, as overlooking terms in the general  equations of motion may lead to ignore important effects these theories may describe.

\section{The equations of motion}
The convention to be employed in the present work is the same as in the Chandrasekhar book \cite{chandra}. In this convention, the signature of the metric tensor $g_{\mu\nu}$ is $(+,-,-,-)$ and the covariant derivative of a generic tensor density $A^{\alpha\beta..}_{\;\;\;\;\;\;\gamma\delta..}$
of weight $w$ is given by
$$
\nabla_\mu A^{\alpha\beta..}_{\;\;\;\;\;\;\gamma\delta..}=\partial_\mu A^{\alpha\beta..}_{\;\;\;\;\;\;\gamma\delta...}+\Gamma^{\alpha}_{\rho\mu}A^{\rho\beta..}_{\;\;\;\;\;\;\gamma\delta..}+\Gamma^\beta_{\rho\mu}A^{\alpha\rho..}_{\;\;\;\;\;\;\gamma\delta..}+...-\Gamma^\rho_{\gamma\mu}A^{\alpha\beta..}_{\;\;\;\;\;\;\rho\delta..}
$$
\be\lb{covdev}
-\Gamma^{\rho}_{\delta\mu}A^{\alpha\beta..}_{\;\;\;\;\;\;\gamma\rho..}-...-w \Gamma^\rho_{\rho \mu} A^{\alpha\beta..}_{\;\;\;\;\;\;\gamma\delta...}.
\ee
The actions to be studied in this letter are typically of the form
\be\lb{action}
S=S_m+\frac{1}{\kappa^2}\int L[g_{\mu\nu}, R^\alpha_{\;\;\;\beta\mu\nu}(\Gamma)]\sqrt{-g}d^4x,
\ee
where the curvature tensor is defined through
\be\lb{curva}
R^\alpha_{\;\;\;\beta\mu\nu}=\partial_{\mu}\Gamma^\alpha_{\beta\nu}-\partial_{\nu}\Gamma^\alpha_{\beta\mu}+\Gamma^\lambda_{\beta\nu}\Gamma^\alpha_{\lambda\mu}-\Gamma^\lambda_{\beta\mu}\Gamma^\alpha_{\lambda\nu},
\ee
the lagrangian $L$ is an arbitrary function of the metric tensor $g_{\mu\nu}$ and  $R^\alpha_{\;\;\;\beta\mu\nu}$, and $S_m$ the matter action. The connection $\Gamma^\alpha_{\beta\gamma}$
is not assumed to be the Levi-Civita one a priori, and it will be considered as an additional variable, independent on the metric $g_{\mu\nu}$. The adequate formalism for dealing with this situation is the Palatini one. In this context,
the unique symmetry property of the curvature that is evident from its definition is $$R^\alpha_{\;\;\;\beta\mu\nu}=-R^\alpha_{\;\;\;\beta\nu\mu}.$$
Instead, the other symmetry properties of the Riemann tensor that are usually considered in  GR  are not necessarily true here.
In the Chandrasekhar convention just introduced, the following identity for the variation of the curvature
\be\lb{varo}
\delta R^\alpha_{\beta\mu\nu}=\nabla_\mu \delta\Gamma^\alpha_{\beta\nu}-\nabla_\nu\delta\Gamma^\alpha_{\beta\mu}-2T^\rho_{\mu\nu}\delta\Gamma^\alpha_{\beta\rho},
\ee
is found \cite{nikodem}. Here the last term involves the torsion tensor of the generalized connection, which is defined by the following expression
$$
T^\alpha_{\beta\gamma}=\frac{1}{2} (\Gamma^\alpha_{\beta\gamma}-\Gamma^\alpha_{\gamma\beta}).
$$
The variation of the action (\ref{action}) with respect to $g_{\mu\nu}$ and $\Gamma_{\alpha\beta}^\gamma$ taken as independent variables is given by
$$
\delta S=\delta S_m+\frac{1}{\kappa^2}\int\sqrt{-g}\bigg[\bigg(\frac{\partial L}{\partial g^{\mu\nu}}-\frac{1}{2}g_{\mu\nu} L\bigg)\delta g^{\mu\nu}+D_\alpha^{\;\;\;\beta\mu\nu}\delta R^\alpha_{\;\;\;\beta\mu\nu}\bigg]d^4x,
$$
where the following quantity 
$$
D_{\alpha}^{\;\;\;\beta\mu\nu}=\frac{\partial L}{\partial R^\alpha_{\beta\mu\nu}},
$$
has been introduced, which is anti-symmetric with respect of  an interchange between $\mu$ and $\nu$.
By taking this symmetry property into account, and by use of (\ref{varo}), the last variation may be expressed as follows
$$
\delta S=\delta S_m+\frac{1}{\kappa^2}\int\sqrt{-g}\bigg[\bigg(\frac{\partial L}{\partial g^{\mu\nu}}-\frac{1}{2}g_{\mu\nu} L\bigg)\delta g^{\mu\nu}+D_\alpha^{\;\;\;\beta\mu\nu}(\nabla_\mu \delta\Gamma^\alpha_{\beta\nu}-T^\rho_{\mu\nu}\delta\Gamma^\alpha_{\beta\rho})\bigg]d^4x,
$$
$$
=\delta S_m+\frac{1}{\kappa^2}\int\bigg[\sqrt{-g}\bigg(\frac{\partial L}{\partial g^{\mu\nu}}-\frac{1}{2}g_{\mu\nu} L\bigg)\delta g^{\mu\nu}+\nabla_\mu(P_\alpha^{\;\;\;\beta\mu\nu}\delta\Gamma^\alpha_{\beta\nu})-\nabla_\mu(P_\alpha^{\;\;\;\beta\mu\nu})\delta\Gamma^\alpha_{\beta\nu}-P_\alpha^{\;\;\;\beta\mu\lambda}T^\nu_{\mu\lambda}\delta\Gamma^\alpha_{\beta\nu}\bigg]d^4x.
$$
In the last expression
$$
P_{\alpha}^{\beta\mu\nu}=\sqrt{-g}\;D_\alpha^{\;\;\;\beta\mu\nu},
$$
which is a tensor density of weight $w=1$ due to the presence of the metric determinant. From  (\ref{covdev}) applied to the case $w=1$ it is found that
$$
\nabla_\mu(P_{\alpha}^{\;\;\;\beta\mu\nu} \delta \Gamma^\alpha_{\beta\nu})=\partial_\mu(P_{\alpha}^{\;\;\;\beta\mu\nu} \delta \Gamma^\alpha_{\beta\nu})+2T^\mu_{\rho\mu} P_{\alpha}^{\;\;\;\beta\rho\nu}\delta \Gamma^\alpha_{\beta\nu}.
$$
By help of this formula, the last action variation may be expressed like this
$$
\delta S=\delta S_m+\frac{1}{\kappa^2}\int\bigg[\sqrt{-g}\bigg(\frac{\partial L}{\partial g^{\mu\nu}}-\frac{1}{2}g_{\mu\nu} L\bigg)\delta g^{\mu\nu}+\partial_\mu(P_\alpha^{\;\;\;\beta\mu\nu}\delta\Gamma^\alpha_{\beta\nu})-\nabla_\mu(P_\alpha^{\;\;\;\beta\mu\nu})\delta\Gamma^\alpha_{\beta\nu}
$$
$$
-P_\alpha^{\;\;\;\beta\mu\lambda}T^\nu_{\mu\lambda}\delta\Gamma^\alpha_{\beta\nu}+2 T^\rho_{\mu\rho} P_{\alpha}^{\;\;\;\beta\mu\nu} \delta\Gamma^\alpha_{\beta\nu}\bigg]d^4x.
$$
By throwing a total derivative in $\delta S$, the following equations of motion are found
\be\lb{in}
\nabla_\mu P_\alpha^{\beta\mu\nu}=-T^\nu_{\sigma\rho}P_\alpha^{\;\;\;\beta\sigma\rho}-2T_{\rho\mu}^\rho P_\alpha^{\;\;\;\beta\mu\nu},
\ee
\be\lb{in2}
\frac{\partial L}{\partial g_{\mu\nu}}-\frac{L}{2}g_{\mu\nu}=\kappa^2 T_{\mu\nu}.
\ee
For a generic curvature tensor $R^\mu_{\alpha\nu\beta}$ there are two possible definitions of a Ricci tensor, which are given by
\be\lb{ricci1}
\widetilde{R}_{\mu\nu}=R^\rho_{\mu\rho\nu},
\ee
\be\lb{ricci2}
R_{\mu}^\nu=g^{\alpha\beta}R^\mu_{\alpha\nu\beta}.
\ee
The theories whose lagrangian is a function of (\ref{ricci1}) have been considered in \cite{olmo}. The purpose of this note is to consider the other possibility namely,
that $L$ is a function of $R_{\mu}^\nu$ solely. The tensor  $P_\alpha^{\beta\mu\nu}$ in this case
is given by the following expression
$$
P_\alpha^{\beta\mu\nu}=\sqrt{-g}W_\alpha^\nu g^{\beta\mu}-\sqrt{-g}W_\alpha ^\mu g^{\beta\nu},
$$
with
\be\lb{wdef}
W_\mu^\nu=\frac{1}{4}\frac{\partial L}{\partial R^{\mu}_\nu}.
\ee
The quantity $P_\alpha^{\beta\mu\nu}$ is a pseudo-tensor density of weight $w=1$ as well.
Then (\ref{covdev}) implies, for any generic connection, that 
$$
\nabla_\mu[\sqrt{-g}W_\alpha^\nu g^{\beta\gamma}]=\partial_\mu[\sqrt{-g}W_\alpha^\nu g^{\beta\gamma}]
+\Gamma^\nu_{\rho\mu} \sqrt{-g}W_\alpha^\rho g^{\beta\gamma}
+\Gamma^\beta_{\rho\mu} \sqrt{-g}W_\alpha^\nu g^{\rho\gamma}
$$
$$
+\Gamma^\gamma_{\rho\mu} \sqrt{-g}W_\alpha^\nu g^{\beta\rho}-\Gamma^\rho_{\alpha\mu} \sqrt{-g}W_\rho^\nu g^{\beta\gamma}-\Gamma^\rho_{\rho\mu} \sqrt{-g}W_\alpha^\nu g^{\beta\gamma}.
$$
In these terms, the equation (\ref{in}) becomes
$$
\nabla^l_\mu[\sqrt{-g}W_\alpha^\nu g^{\beta\mu}]-\nabla^l_\mu[\sqrt{-g}W_\alpha^\mu g^{\beta\nu}]
+\Delta\Gamma^\nu_{\rho\mu} \sqrt{-g}W_\alpha^\rho g^{\beta\mu}
+\Delta\Gamma^\beta_{\rho\mu} \sqrt{-g}W_\alpha^\nu g^{\rho\mu}
$$
$$
+\Delta\Gamma^\mu_{\rho\mu} \sqrt{-g}W_\alpha^\nu g^{\beta\rho}-\Delta\Gamma^\rho_{\alpha\mu} \sqrt{-g}W_\rho^\nu g^{\beta\mu}-\Delta\Gamma^\rho_{\rho\mu} \sqrt{-g}W_\alpha^\nu g^{\beta\mu}
 -\Delta\Gamma^\mu_{\rho\mu} \sqrt{-g}W_\alpha^\rho g^{\beta\nu}
 $$
 $$
-\Delta\Gamma^\beta_{\rho\mu} \sqrt{-g}W_\alpha^\mu g^{\rho\nu}
-\Delta\Gamma^\nu_{\rho\mu} \sqrt{-g}W_\alpha^\mu g^{\beta\rho}
+\Delta\Gamma^\rho_{\alpha\mu} \sqrt{-g}W_\rho^\mu g^{\beta\nu}+\Delta\Gamma^\rho_{\rho\mu} \sqrt{-g}W_\alpha^\mu g^{\beta\nu}
$$
$$
=-T^\nu_{\sigma\rho}(\sqrt{-g}W_\alpha^\rho g^{\beta\sigma}-\sqrt{-g}W_\alpha^\sigma g^{\beta\rho})-2T_{\rho\mu}^\rho(\sqrt{-g}W_\alpha^\nu g^{\beta\mu}-\sqrt{-g}W_\alpha ^\mu g^{\beta\nu}),
$$
with $\nabla^l$ the standard Levi-Civita connection. The full covariant derivative was decomposed above as $\nabla=\nabla^l+\Delta \Gamma$, where $\Delta \Gamma_{\alpha\beta}^\gamma=C_{\alpha\beta}^\gamma+T_{\alpha\beta}^\gamma$, with $C_{\alpha\beta}^\gamma$
the symmetric and $T_{\alpha\beta}^\gamma$ the anti-symmetric parts of the extra components of the connection, respectively\footnote{The reason for using the notation $\Delta \Gamma^\gamma_{\alpha\beta}$ for these quantities is to avoid confusion with the quantities $\delta \Gamma^\gamma_{\alpha\beta}$ employed in the variation of the action $S$ given above. Both quantities are of course different.}. The last expression may be written in terms of these components as 
$$
\nabla^l_\mu[W_\alpha^\nu g^{\beta\mu}]-\nabla^l_\mu[W_\alpha^\mu g^{\beta\nu}]
+C^\nu_{\rho\mu} W_\alpha^\rho g^{\beta\mu}
+C^\beta_{\rho\mu} W_\alpha^\nu g^{\rho\mu}
+C^\mu_{\rho\mu} W_\alpha^\nu g^{\beta\rho}-C^\rho_{\alpha\mu} W_\rho^\nu g^{\beta\mu}-C^\rho_{\rho\mu} W_\alpha^\nu g^{\beta\mu}
 $$
 $$
 -C^\mu_{\rho\mu} W_\alpha^\rho g^{\beta\nu}
-C^\beta_{\rho\mu} W_\alpha^\mu g^{\rho\nu}
-C^\nu_{\rho\mu} W_\alpha^\mu g^{\beta\rho}
+C^\rho_{\alpha\mu} W_\rho^\mu g^{\beta\nu}+C^\rho_{\rho\mu} W_\alpha^\mu g^{\beta\nu}
=-T^\nu_{\sigma\rho}(W_\alpha^\rho g^{\beta\sigma}-W_\alpha^\sigma g^{\beta\rho})
$$
$$
-2T_{\rho\mu}^\rho(W_\alpha^\nu g^{\beta\mu}-W_\alpha ^\mu g^{\beta\nu})
-T^\nu_{\rho\mu} W_\alpha^\rho g^{\beta\mu}
-T^\beta_{\rho\mu} W_\alpha^\nu g^{\rho\mu}
-T^\mu_{\rho\mu} W_\alpha^\nu g^{\beta\rho}+T^\rho_{\alpha\mu} W_\rho^\nu g^{\beta\mu}+T^\rho_{\rho\mu} W_\alpha^\nu g^{\beta\mu}
 $$
 $$
 +T^\mu_{\rho\mu} W_\alpha^\rho g^{\beta\nu}
+T^\beta_{\rho\mu} W_\alpha^\mu g^{\rho\nu}
+T^\nu_{\rho\mu} W_\alpha^\mu g^{\beta\rho}
-T^\rho_{\alpha\mu} W_\rho^\mu g^{\beta\nu}-T^\rho_{\rho\mu} W_\alpha^\mu g^{\beta\nu}.
$$
Here the fact that $\nabla^l\sqrt{-g}=0$ was employed to cancel the metric determinant everywhere, and all the torsion terms were moved to the right hand side. On the left hand side the fifth and the seventh term cancel each other, and the same holds for the eight and the twelfth and the third and the tenth. On the right hand side the seventh and the nineth term are the same, and the same holds for the tenth and the last one and the fifth and the twelfth. 
Therefore the last expression may be written more concisely as
$$
\nabla^l_\mu[W_\alpha^\nu g^{\beta\mu}]-\nabla^l_\mu[W_\alpha^\mu g^{\beta\nu}]
+C^\beta_{\rho\mu} (W_\alpha^\nu g^{\rho\mu}-W_\alpha^\mu g^{\rho\nu})
+C^\rho_{\alpha\mu}(W_\rho^\mu g^{\beta\nu}-W_\rho^\nu g^{\beta\mu})
$$
$$
=-T^\nu_{\sigma\rho}(W_\alpha^\rho g^{\beta\sigma}-W_\alpha^\sigma g^{\beta\rho})
-2T_{\rho\mu}^\rho(W_\alpha^\nu g^{\beta\mu}-W_\alpha ^\mu g^{\beta\nu})
-2T^\nu_{\rho\mu} W_\alpha^\rho g^{\beta\mu}
$$
$$
-2T^\mu_{\rho\mu}(W_\alpha^\nu g^{\beta\rho}
 -W_\alpha^\rho g^{\beta\nu})
 -T^\beta_{\rho\mu}(W_\alpha^\nu g^{\rho\mu}-W_\alpha^\mu g^{\rho\nu})
-T^\rho_{\alpha\mu}(W_\rho^\mu g^{\beta\nu}-W_\rho^\nu g^{\beta\mu}).
$$
Further simplifications are possible. The right hand the first term is of the same type as fifth, up to a sign and a factor 2. The third and fourth cancel the sixth and seventh.
By making the corresponding simplifications, this leads to 
$$
\nabla^l_\mu[W_\alpha^\nu g^{\beta\mu}]-\nabla^l_\mu[W_\alpha^\mu g^{\beta\nu}]
+C^\beta_{\rho\mu} [W_\alpha^\nu g^{\rho\mu}-W_\alpha^\mu g^{\rho\nu}]
+C^\rho_{\alpha\mu}[W_\rho^\mu g^{\beta\nu}-W_\rho^\nu g^{\beta\mu}]
$$
$$
= T^\beta_{\rho\mu}[W_\alpha^\mu g^{\rho\nu}-W_\alpha^\nu g^{\rho\mu}]
-T^\rho_{\alpha\mu}[ W_\rho^\mu g^{\beta\nu}-W_\rho^\nu g^{\beta\mu}].
$$
Note that the second term on the right hand side is zero, as is a product of a symmetric and anti-symmetric quantities. It was added for convenience, in order to show that the last formula can be written in very simple fashion as 
\be\lb{simfash}
\nabla^l_\mu[W_\alpha^\nu g^{\beta\mu}]-\nabla^l_\mu[W_\alpha^\mu g^{\beta\nu}]
+\Delta\Gamma^\beta_{\rho\mu} [W_\alpha^\nu g^{\rho\mu}-W_\alpha^\mu g^{\rho\nu}]
+\Delta\Gamma^\rho_{\alpha\mu}[W_\rho^\mu g^{\beta\nu}-W_\rho^\nu g^{\beta\mu}]=0.
\ee
It may be desirable to eliminate the term $\nabla^l_\mu W_\alpha^\mu$. This is achieved by multiplying the last identity by $g_{\beta\nu}$, which throws the following result
$$
\nabla^l_\mu W_\alpha^\mu=\frac{1}{3}g^{\rho\mu}\Delta\Gamma^\sigma_{\rho\mu} W_{\alpha\sigma} -\frac{1}{3}\Delta\Gamma^\sigma_{\sigma\mu}W_\alpha^\mu
+\Delta\Gamma^\rho_{\alpha\mu}W_\rho^\mu.
$$
The insertion of the last formula into the equation (\ref{simfash}) leads to
$$
\nabla^l_\mu [W_\alpha^\nu g^{\beta\mu}]+\Delta\Gamma^\beta_{\rho\mu} [W_\alpha^\nu g^{\rho\mu}-W_\alpha^\mu g^{\rho\nu}]
+\Delta\Gamma^\rho_{\alpha\mu}[W_\rho^\mu g^{\beta\nu}-W_\rho^\nu g^{\beta\mu}]
$$
$$
-\frac{1}{3} g^{\beta\nu}g^{\rho\mu}\Delta\Gamma^\sigma_{\rho\mu}W_{\alpha\sigma} +\frac{1}{3} g^{\beta\nu}\Delta\Gamma^\sigma_{\sigma\mu} W_\alpha^\mu
-g^{\beta\nu}\Delta\Gamma^\rho_{\alpha\mu}W_\rho^\mu=0.
$$
The fourth and the last term cancel each other, thus the last expression can be simplified to 
$$
\nabla^l_\mu [W_\alpha^\nu g^{\beta\mu}] +\Delta\Gamma^\beta_{\rho\mu} [W_\alpha^\nu g^{\rho\mu}-W_\alpha^\mu g^{\rho\nu}]
-\Delta\Gamma^\rho_{\alpha\mu}W_\rho^\nu g^{\beta\mu}
-\frac{1}{3} g^{\beta\nu}g^{\rho\mu}\Delta\Gamma^\sigma_{\rho\mu} W_{\alpha\sigma} +\frac{1}{3} g^{\beta\nu}\Delta\Gamma^\sigma_{\sigma\mu} W_\alpha^\mu=0.
$$
By multiplying the last identity by $g_{\beta\delta}$ and  $g_{\nu\eta}$ it is found, after renaming  some indices, that
\be\lb{endlich}
\nabla^l_\mu W_{\alpha\eta} =-g_{\beta\mu}g^{\rho\epsilon}\Delta\Gamma^\beta_{\rho\epsilon} W_{\alpha\eta} +\Delta\Gamma^\rho_{\alpha\mu}W_{\rho\eta}
+[g_{\beta\mu}g^{\epsilon\rho} \Delta\Gamma^\beta_{\eta\epsilon}
+\frac{1}{3} g_{\eta\mu}g^{\sigma\epsilon}\Delta\Gamma^\rho_{\sigma\epsilon} -\frac{1}{3} g_{\eta\mu}g^{\rho\epsilon}\Delta\Gamma^\sigma_{\sigma\epsilon}]W_{\alpha\rho}.
\ee
Thus, the  equations of motion (\ref{in}) has been reduced to (\ref{endlich}). These equations are to be complemented with the equation of motion (\ref{in2}).

The next step is to consider some situations in which $W_{\mu\nu}$ has additional symmetries, which allows to simplify further the system obtained above.
\section{Some particular cases}
\subsection{The case where $W_{\mu\nu}$ is symmetric}
Assume now that $W_{\mu\nu}=W_{\nu\mu}$. This may be the case, for instance, when the lagragian is such that $L=L(R_{(\mu\nu)})$.
In this situation, it follows that  
$\nabla_\gamma W_{\alpha\eta}=\nabla_{\gamma}W_{\eta\alpha}$, as the difference is the covariant derivative of the zero tensor. By inspection  of (\ref{endlich}) it is seen that
this condition implies that
\be\lb{constr2}
\Delta\Gamma^\rho_{\eta\mu}=g_{\beta\mu}g^{\epsilon\rho} \Delta\Gamma^\beta_{\eta\epsilon}
+\frac{1}{3} g_{\eta\mu}g^{\sigma\epsilon}\Delta\Gamma^\rho_{\sigma\epsilon} -\frac{1}{3} g_{\eta\mu}g^{\rho\epsilon}\Delta\Gamma^\sigma_{\sigma\epsilon}.
\ee
This is a constraint involving $C_{\alpha\beta}^\gamma$ and $T_{\alpha\beta}^\gamma$. The contraction of $\rho$ and $\eta$ give
\be\lb{constr3}
\Delta\Gamma^\rho_{\rho\mu}=g_{\rho\mu}g^{\epsilon\sigma} \Delta\Gamma^\rho_{\sigma\epsilon}.
\ee
The contraction of $\rho$ and $\mu$ together with (\ref{constr3}) result into a trivial identity.
When the last formula (\ref{constr3}) is inserted back in (\ref{constr2}), it is deduced that
$$
\Delta\Gamma^\rho_{\eta\mu}=g_{\beta\mu}g^{\epsilon\rho} \Delta\Gamma^\beta_{\eta\epsilon}.
$$
By defining $\Delta\Gamma_{\lambda\eta\mu}=g_{\rho\lambda}\Delta\Gamma^\rho_{\eta\mu}$ the last condition implies that
\be\lb{constr}
\Delta\Gamma_{\lambda\eta\mu}= \Delta\Gamma_{\mu\eta\lambda}.
\ee
This formula has important consequences. When expressed in terms of $C^\rho_{\mu\nu}$ and $T^\rho_{\mu\nu}$
it gives
\be\lb{constrai}
C_{\lambda\eta\mu}-C_{\mu\eta\lambda}=T_{\mu\eta\lambda}-T_{\lambda\eta\mu}.
\ee
By making cyclic permutations of the indices of this formula and adding and subtracting the resulting identities conveniently, it may be shown that
\be\lb{tordes}
T_{\mu\nu\alpha}=C_{\alpha\mu\nu}-C_{\nu\mu\alpha}.
\ee
This means that, once the symmetric part of the extra part of the connection $C_{\alpha\mu\nu}$ has been calculated, the the torsion is automatically found.
Now, when the constraints (\ref{constr2})-(\ref{constr}) are inserted into (\ref{endlich}),  the resulting expression is found
$$
\nabla^l_\mu W_{\alpha\eta}=-g_{\beta\mu}g^{\rho\epsilon}\Delta\Gamma^\beta_{\rho\epsilon} W_{\alpha\eta} +\Delta\Gamma^\rho_{\alpha\mu}W_{\rho\eta}
+\Delta\Gamma^\rho_{\eta\mu}W_{\alpha\rho}.
$$
This can be written,  by use of (\ref{constr3}), as 
$$
\nabla^l_\mu W_{\alpha\eta}=- \Delta\Gamma^\rho_{\rho\mu} W_{\alpha\eta} +\Delta\Gamma^\rho_{\alpha\mu}W_{\rho\eta}
+\Delta\Gamma^\rho_{\eta\mu}W_{\alpha\rho}.
$$
 By multiplying this expression by the inverse $(W^{-1})^{\alpha\eta}$ of $W_{\alpha\eta}$, and by taking into account the well known formula of the
derivative of the logarithm of a determinant together with the Levi-Civita contraction of the Christoffel symbol $\Gamma^{(l)\rho}_{\mu\rho}=\partial_\mu \log \sqrt{-g}$,
 the following identity is found
\be\lb{cobos2}
\partial_\mu \log \bigg(-\frac{W}{g}\bigg)=-2\Delta\Gamma^\alpha_{\alpha\mu}.
\ee
Thus
$$
\nabla^l_\mu W_{\alpha\eta}=\bigg(\partial_\mu \log \sqrt{-\frac{W}{g}} \bigg)W_{\alpha\eta} +\Delta\Gamma^\rho_{\alpha\mu}W_{\rho\eta}
+\Delta\Gamma^\rho_{\eta\mu}W_{\alpha\rho}
$$
The first term in the right hand side of the last expression may be erased by making the following conformal like transformation
\be\lb{zetadef}
Z_{\alpha\beta}=\sqrt{-\frac{g}{W}}W_{\alpha\beta}.
\ee
In other words $Z_{\alpha\beta}$ is a conformal scaling of $W_{\alpha\beta}$ in such a way that $|Z|=|g|$.
When the last equation is expressed in terms of $Z_{\alpha\beta}$ the result is
\be\lb{cobos}
 \nabla^l_\mu Z_{\alpha\eta}=\Delta\Gamma^\rho_{\alpha\mu} Z_{\rho\eta}
+\Delta\Gamma^\rho_{\eta\mu} Z_{\alpha\rho}.
\ee
This means that, if $Z_{\mu\nu}$ is interpreted as a metric, then the full connection of the theory is metric compatible
\be\lb{cobos2}
\nabla_\mu Z_{\alpha\beta}=0.
\ee
This fact makes the description of the resulting theory less complicated.
In fact, by making cyclic permutations of (\ref{cobos2}) and adding and subtracting them conveniently, it is concluded that
\be\lb{despejest}
\partial_\alpha Z_{\mu\nu}+\partial_\nu Z_{\alpha\mu}-\partial_\mu Z_{\nu\alpha}=2 T_{\mu\alpha}^\rho Z_{\nu\rho}+2( \Gamma_{\nu\alpha}^{(l)\rho}+C_{\nu\alpha}^\rho )Z_{\rho\mu}+2 T_{\mu\nu}^\rho Z_{\rho\alpha},
\ee
with $\Gamma_{\nu\alpha}^{(l)\beta}$ the standard Christoffel symbols for the Levi-Civita connection. This gives
\be\lb{gibbs}
\Gamma_{\nu\alpha}^{(l)\beta}+C_{\nu\alpha}^\beta =\frac{1}{2}(Z^{-1})^{\beta\mu}(\partial_\alpha Z_{\mu\nu}+\partial_\nu Z_{\alpha\mu}-\partial_\mu Z_{\nu\alpha})+ T_{\alpha\mu}^\rho Z_{\nu\rho}(Z^{-1})^{\beta\mu}+ T_{\nu\mu}^\rho Z_{\rho\alpha}(Z^{-1})^{\beta\mu}.
\ee
From here, it is seen that the symmetric part of the connection can be expressed in terms of the torsion by the formula
$$
C_{\nu\alpha}^\beta =\frac{1}{2}(Z^{-1})^{\beta\mu}(\partial_\alpha Z_{\mu\nu}+\partial_\nu Z_{\alpha\mu}-\partial_\mu Z_{\nu\alpha})-\frac{1}{2}g^{\beta\mu}(\partial_\alpha g_{\mu\nu}+\partial_\nu g_{\alpha\mu}-\partial_\mu g_{\nu\alpha})
$$
\be\lb{tumuch}
+ T_{\alpha\mu}^\rho Z_{\nu\rho}(Z^{-1})^{\beta\mu}+ T_{\nu\mu}^\rho Z_{\rho\alpha}(Z^{-1})^{\beta\mu}.
\ee
On the other hand, (\ref{tordes}) and  the last formula together imply that
$$
T_{ \nu\alpha\eta} - T_{\rho\alpha\mu} Z^\rho_{\nu}(Z^{-1})^{\mu}_\eta-T_{\rho\nu\mu} Z^\rho_{\alpha}(Z^{-1})^{\mu}_\eta+T_{\rho\eta\mu} Z^\rho_{\nu}(Z^{-1})^{\mu}_\alpha+T_{\rho\nu\mu} Z^\rho_{\eta}(Z^{-1})^{\mu}_\alpha
$$
$$
=\frac{1}{2}(Z^{-1})^{\mu}_\eta (\partial_\alpha Z_{\mu\nu}+\partial_\nu Z_{\alpha\mu}-\partial_\mu Z_{\nu\alpha})
-\frac{1}{2}(Z^{-1})^{\mu}_\alpha (\partial_\eta Z_{\mu\nu}+\partial_\nu Z_{\eta\mu}-\partial_\mu Z_{\nu\eta})
$$
\be\lb{linear}
+\partial_\eta g_{\alpha\nu}-\partial_\alpha g_{\nu\eta}.
\ee
In four dimensions, the tensor $T_{ \nu\alpha\eta}$ has 24 independent components. Therefore, the last equation is a linear system for finding  $T_{ \nu\alpha\eta}$, which involves a $24\times 24$ matrix coefficients
constructed in terms of products of monomials of the form $Z^\rho_{\eta}(Z^{-1})^{\beta}_\alpha$. If it has a solution, it would be expressed in terms of $\partial_\alpha Z_{\beta\gamma}$,  $\partial_\alpha g_{\beta\gamma}$ and these monomials. It is cumbersome find the inverse of a  $24\times 24$ matrix, but it is important to understand  if this linear system admits some sort of solutions, or if it is non compatible. It is also important to understand if the torsion can be imposed to zero or if this will lead to an overdetermined system. 

Before to finish this section, it should be remarked that for the Einstein Hilbert lagrangian $L=(8\pi G_N)^{-1} R$ it is easily deduced from (\ref{zetadef}) that $Z_{\mu\nu}=\lambda g_{\mu\nu}$. In this particular case one may impose the condition of zero torsion. Then the condition $\nabla Z_{\mu\nu}=0$ implies that the connection is the Levi-Civita one for $g_{\mu\nu}$ and that the equations of motion (\ref{in2}) become the standard Einstein equations
$$
R_{\mu\nu}-\frac{g_{\mu\nu}}{2}R=G_N T_{\mu\nu}.
$$
This is important, as the Einstein equations are known to arise in both the metric and Palatini formalism, when the condition of zero torsion is imposed.  There exist other models for which the zero torsion condition may be imposed, for instance in Lovelock theories \cite{borunda2}. However, for $f(R)$, this results in a more severe restriction and only holds for maximally symmetrical space times \cite{borunda}. For other models, the absence of torsion may be problematic, and may lead to no solutions at all. The next section is devoted to discuss this important issue.

\section{The difficulties of imposing the zero torsion condition}
In the present section it will be argued that it is not possible in general to impose the zero torsion condition, for the models in consideration. All these statements are valid of course for symmetric $W_{\mu\nu}$. These affirmations will be suggested by simplicity by a $1+1$ dimensional example, and  the resulting arguments will
be generalized later on to higher dimensions. But before to start this discussion, it is perhaps convenient to consider the following proposition.
\\

\textbf{Proposition:} If the quantity $Z_{\mu\nu}$ defined in (\ref{zetadef}) is such that $Z_{\mu\nu}=\lambda g_{\mu\nu}$, the the unique connection with torsion satisfying  $\nabla g_{\mu\nu}=0$ is the Levi-Civita one. 
\\

\texttt{Proof:} In a general model, this proposition would be false. However, the proposition is true due to the condition (\ref{constr}), which is valid for symmetric $W_{\mu\nu}$. In effect, this condition implies that $\Delta\Gamma_{\alpha\beta\gamma}$ is symmetric with the interchange of the first and third indices. On the other hand the condition $\nabla g_{\mu\nu}=0$
can be written as
$$
\nabla_\alpha^l g_{\mu\nu}+\Delta\Gamma^{\rho}_{\mu\alpha} g_{\rho\nu}+\Delta\Gamma^{\rho}_{\nu\alpha} g_{\mu\rho}=0
$$
This implies that
$$
\Delta\Gamma_{\alpha\beta\gamma}=-\Delta\Gamma_{\beta\alpha\gamma}
$$
thus this quantity is antisymmetric with the interchange of the first and second indices. This, together with the other symmetry property implies that
$$
\Delta\Gamma_{\alpha\beta\gamma}=-\Delta\Gamma_{\gamma\alpha\beta},
$$
thus, this quantity changes the sign under a cyclic permutation. However, this implies that
$$
\Delta\Gamma_{\alpha\beta\gamma}=-\Delta\Gamma_{\gamma\alpha\beta}=\Delta\Gamma_{\beta\gamma\alpha}=-\Delta\Gamma_{\alpha\beta\gamma}.
$$
The first and the last term of this identity show that $\Delta\Gamma_{\alpha\beta\gamma}=0$, thus the connection is simply the Levi-Civita one. $\square$
\\

\textbf{Remark 1:} If the non quantity $C^\alpha_{\mu\nu}$ is zero, then the torsion $C^\alpha_{\mu\nu}$ vanishes. 
\\

This is easily seen by the equation (\ref{tordes}).$\square$
\\

\textbf{Remark 2:} If the torsion $T^\alpha_{\mu\nu}$ is zero, then the quantity $C_{\alpha\mu\nu}$ is symmetric with respect to the interchange of  any pair of indices. 
\\

This is easily seen by the equation (\ref{constrai}), together with the symmetry property involving the second and third indices.$\square$
\\

The possibility described in remark 2 will be considered below, and it will be argued that is not realized in general models.

\subsection{The case of zero torsion but non zero $C_{\mu\nu\alpha}$}

In the following, it is studied the possibility of imposing the torsion to zero but allowing the quantity 
$C^\mu_{\nu\alpha}$ to be non trivial. It is argued however, that it is likely that there are no solutions, except in very specific cases.

In order to see the complications for finding such solutions, it is of interest to consider a two dimensional model.
Recall that the choice of curvature considered here is given by
\be\lb{elijo}
R_{\mu}^\nu=g^{\alpha\beta}R^\mu_{\alpha\nu\beta},\qquad R=R^\mu_\mu.
\ee
There are some subtle issues to be considered before proceed. In two dimensions, for a generic metric
$g_{\mu\nu}$, the following simple formulas are valid in the Levi-Civita case
\be\lb{tudim}
R^\mu_{\alpha\nu\beta}=\delta^\mu_{[\nu}g_{\beta]\alpha} R,\qquad
R^\mu_{\nu}=\frac{1}{2}\delta^\mu_{\nu} R.
\ee
In addition, the scalar curvature is given by
\be\lb{tudim2}
R=\frac{2R_{1212}}{|g|}.
\ee
On the other hand, the assumption that $C^\mu_{\nu\alpha}$ is non zero but $T^\mu_{\nu\alpha}=0$, together with (\ref{cobos2}) implies that, for the present case, the tensor
$R^\mu_{\alpha\nu\beta}$ corresponds to the Levi-Civita one constructed in terms of the metric $Z_{\mu\nu}$. However, this do not imply that (\ref{tudim})-(\ref{tudim2}) are the appropriate ones for dealing with the present case.
The problem is that the definitions (\ref{elijo}) involve, in addition to $Z_{\mu\nu}$,  the physical metric $g_{\mu\nu}$. Nevertheless, the correct formulas can be found without difficulty.
In order to find these formulas, the quantities constructed entirely in terms of $Z_{\mu\nu}$ will be denoted with a bar, for instance, the curvature tensor is denoted as $\bar{R}^\mu_{\alpha\nu\beta}$
and so on. Note that, even taking into account the subtlety described above, the first formula (\ref{tudim}) is valid if $Z_{\mu\nu}$ is taken as the metric. In other words
$$
\overline{R}^\mu_{\alpha\nu\beta}=\delta^\mu_{[\nu}Z_{\beta]\alpha} \overline{R}
$$
In addition, one has that $\overline{R}^\mu_{\alpha\nu\beta}=R^\mu_{\alpha\nu\beta}$, since its definition at this point does not involve $g_{\mu\nu}$. By taking this into account, and by multiplying
the last expression by the metric tensor $g^{\alpha\beta}$, it is found that 
\be\lb{tudim3}
R^\mu_{\nu}=\delta^\mu_{[\nu}Z_{\beta]}^\beta \overline{R},\qquad R=\frac{1}{2}Z^\alpha_\alpha \overline{R}
\ee
It should be emphasized that neither the quantities with bar or without bars necessarily correspond to the Levi-Civita case constructed entirely in terms of the physical metric $g_{\mu\nu}$. 
The formulas (\ref{tudim3}) are the adequate for studying two dimensional solutions, as it will be done below.

Equipped with the machinery above, consider a generic two dimensional lagrangian. For fixing ideas the following lagrangian
$$
L=\alpha R_{(\mu\nu)}R^{(\mu\nu)},
$$
will be considered, but there is nothing essential in this choice and can be replaced by other ones as well. 
The equations of motion corresponding to this lagrangian are
\be\lb{turco}
2\alpha R_{(\mu\nu)} R_{(\alpha\beta)} g^{\alpha\mu}-\frac{\alpha}{2}g_{\beta\nu}R_{(\alpha\mu)}R^{(\alpha\mu)}=T_{\mu\nu}.
\ee
By taking the trace it is found that
\be\lb{traza}
\alpha R_{(\mu\nu)}R^{(\mu\nu)}=T.
\ee
Thus, the equations of motion (\ref{turco}) may be written as
\be\lb{turco2}
2\alpha R_{(\mu\nu)} R_{(\alpha\beta)} g^{\alpha\mu}=T_{\mu\nu}+\frac{\alpha}{2}g_{\beta\nu}T.
\ee
On the other hand, the formulas (\ref{tudim3}) shows that the last expression is equivalent to
\be\lb{turco3}
\frac{\alpha}{2}[g_{\beta\nu} (Z_{\gamma}^{\gamma})^2-2 Z_{\beta\nu} Z_{\gamma}^{\gamma}+Z_{\alpha\beta} Z_{\nu}^{\alpha}] \overline{R}^2=T_{\mu\nu}+\frac{1}{2}g_{\beta\nu}T.
\ee
An interesting task may be to solve the equations (\ref{turco3}) for an homogeneous and isotropic model. The matter fields sourcing the tensor $T^{\mu\nu}$ are assumed to depend only on the proper time $t$. Usually, the equations of motion of the matter field depends on the metric, as for instance for an scalar field or dust or radiation. Therefore one may assume that the metric has the standard form
$$
g=dt^2-a^2(t)dx^2.
$$
If this were not the case, an explicit dependence on the coordinate $x$ may appear for the matter fields, contradicting the previous hypothesis. It is also likely that the
tensor $Z_{\mu\nu}$ is  time dependent only. This can be seen by assuming an explicit $x$ dependence of $Z_{\mu\nu}$ in (\ref{turco3}), which implies that 
$$
\overline{R}=\frac{f_{\beta\nu}(t)}{g_{\beta\nu} (Z_{\gamma}^{\gamma})^2-2 Z_{\beta\nu} Z_{\gamma}^{\gamma}+Z_{\alpha\beta} Z_{\nu}^{\alpha}},
$$
for any choice of indices $\beta\nu$, with $f_{\beta\nu}(t)$ simple proper time functions.
This is needed since the right hand of (\ref{turco3}), which is related to the physical metric and the matter fields, is only time dependent. To require such dependence for $\overline{R}$ regardless the choice of indices, for any matter field content,  is very restrictive. Therefore, it is safer to assume that the metric $Z_{\mu\nu}$ is $x$ independent and has the general dependence
\be\lb{deerfor}
Z=b^2(t) dt^2-d^2(t)dx dt-c^2(t) dx^2,
\ee
with $b(t)$, $c(t)$ and $d(t)$ functions to be determined by the dynamics. This does not imply any particular constraint about $\overline{R}$, since it will be automatically $x$ independent.

The next task is to understand the consequences of the requirement of zero torsion. This will follow from the relations (\ref{tordes}) and (\ref{tumuch}). As the remark 2 shows, this in particular implies that $C_{\mu\nu\alpha}$ is symmetric with respect to the interchange of any indices.
In order to derive the consequences of this observations, one should recall that in two dimensions there are six independent components of the symbols 
$$
\gamma_{\rho\alpha\nu}=\frac{1}{2}(\partial_\alpha Z_{\rho\nu}+\partial_\nu Z_{\alpha\rho}-\partial_\rho Z_{\nu\alpha}),
$$
namely $\gamma_{111}$, $\gamma_{112}$, $\gamma_{211}$, $\gamma_{222}$, $\gamma_{221}$ and $\gamma_{122}$. The equations (\ref{tordes}) and (\ref{tumuch}, for the case for which $T_{\mu\nu}^\alpha$ show then that
$$
-\partial_1 Z_{22}=-\frac{1}{2}Z_{1\beta}g^{\beta\mu}\partial_\mu g_{22}-\frac{1}{2}Z_{2\beta}g^{\beta\mu}(\partial_1 g_{2\mu}-\partial_\mu g_{12}),
$$
$$
-\partial_1 Z_{12}=\frac{1}{2}Z_{1\beta}g^{\beta\mu}(\partial_1 g_{2\mu}-\partial_\mu g_{12})-\frac{1}{2}Z_{2\beta}g^{\beta\mu}(2\partial_1 g_{1\mu}-\partial_\mu g_{11}).
$$
Explicitly, the last equations are
\be\lb{fade}
-2 c\dot{c}=b^2 a \dot{a}-c^2 H,\qquad 2 \dot{d}= -H d.
\ee
This relates $c$ and $d$ to $a$ and $b$, thus one may take $a(t)$ and $b(t)$ to be independent functions. In these terms,  one may employ the equations of motion (\ref{turco}) and the matter field equations to finally determine $a(t)$, $b(t)$, $c(t)$ and the matter fields $\phi^m(t)$ as functions of $t$. For instance, for a relativistic simple fluid the equations (\ref{turco3}), 
the energy conservation and the equation of state are respectively
\be\lb{trivio}
2\alpha d\bigg[\frac{1}{b^2}\frac{d^2\log c^2}{dt^2}+\frac{1}{c^2}\frac{d^2}{dt^2}\bigg(\frac{c^2}{b^2}\bigg)\bigg]^2=0,
\ee
\be\lb{turco4}
\frac{2\alpha c^2}{a^4}\bigg[\frac{1}{b^2}\frac{d^2\log c^2}{dt^2}+\frac{1}{c^2}\frac{d^2}{dt^2}\bigg(\frac{c^2}{b^2}\bigg)\bigg]^2=\frac{3\rho}{2}+\frac{1}{2a^2}p,
\ee
\be\lb{turco5}
-2\alpha a^2 b^2\bigg[\frac{1}{b^2}\frac{d^2\log c^2}{dt^2}+\frac{1}{c^2}\frac{d^2}{dt^2}\bigg(\frac{c^2}{b^2}\bigg)\bigg]^2 =-\frac{3p}{2}-\frac{a^2}{2}\rho.
\ee
\be\lb{turco6}
\dot{\rho}+H(\rho+p)=0,\qquad p=p(\rho).
\ee
The first equation corresponds to the choice of indices $xt$ and implies that $d=0$. This information was included when deriving the remaining equations.
The system (\ref{turco4})-(\ref{turco6}) together with (\ref{fade})
is enough for determining $a(t)$, $b(t)$, $c(t)$ and $\rho(t)$, once the state equation $p=p(\rho)$ is given. 
Thus, at this point, the solution of the model has been determined. The same procedure may apply to other matter fields such as radiation
or scalar fields.

However, this is not the end of the calculation. From the definition of $W_{\mu\nu}$ (\ref{wdef}) it is found that
$$
W_{\mu\nu}=\frac{\alpha}{2}R_{(\mu\nu)},
$$
which, together with (\ref{zetadef}) implies that
$$
Z_{\mu\nu}=\sqrt{-\frac{|g|}{|R_{\mu\nu}|}}R_{\mu\nu}.
$$ 
By use of (\ref{tudim3}) it is obtained that
\be\lb{oblig}
Z_{\mu\nu}=\frac{a(t)(g_{\mu\nu} Z_\gamma^\gamma-Z_{\mu\nu})}{|(g_{\mu\nu} Z_\gamma^\gamma-Z_{\mu\nu})|}.
\ee
These constitute into new relations over the already determined time functions  $a(t)$, $b(t)$ and $c(t)$, and in particular they imply that $|Z_{\mu\nu}|=|g_{\mu\nu}|$.
This equality between the determinants imply the following algebraic relation between the time functions
$$
a^2(t)=b^2(t)c^2(t).
$$
The problem is  that those functions were already determined, and it is not granted that this algebraic relation will hold for a generic choice of the state equation $p(\rho)$ or even, for a generic choice of matter fields. The full relations (\ref{oblig}) are not expected to be satisfied either.  The next step is to interpret the reason for this negative result. Note that this reasoning in particular forbids the curvature to be the standard Levi-Civita one constructed in terms of $g_{\mu\nu}$, as is a particular case of the present discussion (which follows by requiring that $C_{\mu\nu\alpha}$ is zero).

 \subsection{The  case with torsion turned on}

The reason for which the equations of motion of the previous section were overdetermined is pretty clear. The equations of motion and the equations (\ref{oblig}) are mandatory. Instead the equations (\ref{fade}) about zero torsion were imposed by hand.  This gives two constraints and, once they are applied, the two equations (\ref{oblig}) are not satisfied. A simple counting shows that there are two extra equations for the system to be determined, in this two dimensional example. This strongly suggest that to impose the torsion to vanish is a practice to be avoided.

As the torsionless case seems dubious, the equations that arise when the torsion is turned on become of interest. Until now, there is no warrant that a non trivial torsion exists, as it is not clear that the system (\ref{linear}) has non trivial solutions. The useful feature of the $1+1$-dimensional case is that there are only two independent components for the torsion, which are $T_{112}$ and $T_{212}$. This  converts (\ref{linear}) it into a $2\times 2$ linear system, which is an enormous simplification. By taking into account that
$$
Z^{-1}=\frac{1}{Z_1^1 Z_2^2-Z_1^2 Z_2^1}
\left(
\begin{array}{ccc}
  Z_2^2  & -Z_1^2  \\
 -Z_2^1   &   Z_1^1
\end{array}
\right),
$$
the solution of the resulting $2\times 2$ system (\ref{linear}) becomes
$$
T_{112}=\frac{Z^2_1 Z^1_1+ Z^2_{1}Z^{2}_2}{2Z^1_1 Z^2_2-2Z^1_2 Z^2_1}\bigg[\frac{1}{2}(Z^{-1})^{\mu}_2 (2\partial_1 Z_{\mu1}-\partial_\mu Z_{11})
-\frac{1}{2}(Z^{-1})^{\mu}_1(\partial_2 Z_{\mu1}+\partial_1 Z_{2\mu}-\partial_\mu Z_{12})
$$
$$
+\partial_2 g_{11}-\partial_1 g_{12}\bigg]
- \frac{Z^1_2 Z^2_1+Z^2_{2}Z^2_2}{2Z^1_1 Z^2_2-2Z^1_2 Z^2_1}\bigg[\frac{1}{2}(Z^{-1})^{\mu}_1 (2\partial_2 Z_{\mu2}-\partial_\mu Z_{22})
$$
$$
-\frac{1}{2}(Z^{-1})^{\mu}_2 (\partial_1 Z_{\mu2}+\partial_2 Z_{1\mu}-\partial_\mu Z_{21})
+\partial_1 g_{22}-\partial_2 g_{12}\bigg],
$$
$$
T_{212}=-\frac{ Z^1_{1}Z^1_1+Z^1_{2}Z^{2}_1}{2Z^1_1 Z^2_2-2Z^1_2 Z^2_1}\bigg[\frac{1}{2}(Z^{-1})^{\mu}_2 (2\partial_1 Z_{\mu1}-\partial_\mu Z_{11})
-\frac{1}{2}(Z^{-1})^{\mu}_1(\partial_2 Z_{\mu1}+\partial_1 Z_{2\mu}-\partial_\mu Z_{12})
$$
$$
+\partial_2 g_{11}-\partial_1 g_{12}\bigg]
+\frac{Z^1_{2}Z^{1}_1+Z^1_{2}Z^2_2}{2Z^1_1 Z^2_2-2Z^1_2 Z^2_1}\bigg[\frac{1}{2}(Z^{-1})^{\mu}_1 (2\partial_2 Z_{\mu2}-\partial_\mu Z_{22})
$$
\be\lb{thor}
-\frac{1}{2}(Z^{-1})^{\mu}_2 (\partial_1 Z_{\mu2}+\partial_2 Z_{1\mu}-\partial_\mu Z_{21})
+\partial_1 g_{22}-\partial_2 g_{12}\bigg].
\ee
This is a non trivial solution and gives confidence that, for generic space time dimensions, a non trivial torsion is to be expected to appear for the models under study.
But the important point is that it presence avoids the equation (\ref{fade}), and the number of equations will match the number of unknowns for the two dimensional isotropic homogeneous case. The price is to work with a curvature $R^\alpha_{\beta\gamma\delta}$ constructed in terms of the connection 
$$
\Gamma_{\nu\alpha}^\beta=\Gamma_{\nu\alpha}^{(l)\beta}+C_{\nu\alpha}^\beta +T^\beta_{\nu\alpha}=\frac{1}{2}(Z^{-1})^{\beta\mu}(\partial_\alpha Z_{\mu\nu}+\partial_\nu Z_{\alpha\mu}-\partial_\mu Z_{\nu\alpha})+ T_{\alpha\mu}^\rho Z_{\nu\rho}(Z^{-1})^{\beta\mu}
$$
\be\lb{gibbs2}
+ T_{\nu\mu}^\rho Z_{\rho\alpha}(Z^{-1})^{\beta\mu}+T^\beta_{\nu\alpha},
\ee
which is more cumbersome,  due to the dependence shown in (\ref{thor}). The system of equations describing the theory is then
\be\lb{lasec1}
R^\alpha_{\;\;\;\beta\mu\nu}=\partial_{\mu}\Gamma^\alpha_{\nu\beta}-\partial_{\nu}\Gamma^\alpha_{\mu\beta}+\Gamma^\lambda_{\beta\nu}\Gamma^\alpha_{\lambda\mu}-\Gamma^\lambda_{\beta\mu}\Gamma^\alpha_{\lambda\nu},
\ee
together with (\ref{in}), which is quoted below by clarity
\be\lb{lasec2}
\frac{\partial L}{\partial g_{\mu\nu}}-\frac{L}{2}g_{\mu\nu}=\kappa^2 T_{\mu\nu}.
\ee
Here
\be\lb{zetadef2}
Z_{\alpha\beta}=\sqrt{-\frac{g}{W}}W_{\alpha\beta},\qquad W_\mu^\nu=\frac{1}{4}\frac{\partial L}{\partial R^{\mu}_\nu}.
\ee
The system described by (\ref{lasec1})-(\ref{lasec2}) and the matter field equations constitute the full field equations.
The fact that the torsion is turned on may be healthy for the system to be well posed, although the resulting equations are apparently more complex.

\subsection{The argument for non trivial torsion in any dimension}

As discussed above, the presence of torsion may  be helpful for issues related to the existence of solutions. The discussion above suggest the following argument ensuring the presence of a non trivial torsion, in any dimensions. 

Assume that the torsion vanishes.
Then the full connection corresponds to the Christoffel symbols $\Gamma^{Z}_{\mu\nu\alpha}$ constructed in terms of the metric $Z_{\alpha\beta}$, as is directly seen by inspection of (\ref{gibbs}).  The curvature $R^\alpha_{\beta\gamma\delta}$ is constructed in terms of these symbols and the corresponding Ricci tensor $R_{\mu\nu}$ turns out to be symmetric. By taking this into account, equations of motion (\ref{lasec1}) reduce to 
\be\lb{ee2}
R^\beta_{\alpha\mu\nu}=F_{\beta\alpha\mu\nu}(g_{\mu\nu}, \partial_{\alpha}g_{\mu\nu}, \partial_{\alpha}\partial_\beta g_{\mu\nu}, Z_{\mu\nu} ,\partial_{\alpha}Z_{\mu\nu}, \partial_{\alpha}\partial_\beta Z_{\mu\nu}).
\ee
On the other hand, the use of the basic definition of $Z_{\alpha\beta}$ in (\ref{zetadef}) allows to write the last equation in the following schematic form
\be\lb{ee3}
R^\beta_{\alpha\mu\nu}=F_{\beta\alpha\mu\nu}(g_{\mu\nu}, \partial_{\alpha}g_{\mu\nu}, \partial_{\alpha}\partial_\beta g_{\mu\nu}, R_{\mu\nu}, \partial_{\alpha}R_{\mu\nu}, \partial_{\alpha}\partial_\beta R_{\mu\nu}).
\ee
By contracting indices, one may obtain from (\ref{ee3}) a system of equations of the form
$$
R_{\mu\nu}=F_{\mu\nu}(g_{\mu\nu}, \partial_{\alpha}g_{\mu\nu}, \partial_{\alpha}\partial_\beta g_{\mu\nu}, R_{\mu\nu}, \partial_{\alpha}R_{\mu\nu}, \partial_{\alpha}\partial_\beta R_{\mu\nu}),
$$
which determine the generalized Ricci tensor $R_{\mu\nu}$ as a function of the metric $g_{\mu\nu}$. Once $R_{\mu\nu}$ have been found in terms of the $g_{\mu\nu}$, the equations (\ref{ee2}) determine the generalized curvature tensor.

Now that this curvature $R^\alpha_{\beta\mu\nu}$ is characterized, at least implicitly, in terms of $g_{\mu\nu}$, the remaining equations of motion (\ref{lasec2}) take the following form
\be\lb{pleto1}
F(g_{\mu\nu}, R_{\mu\nu})=T_{\mu\nu}^m,
\ee
where $T_{\mu\nu}^m$ is the matter field energy-momentum tensor. These equation of motions are to be supplemented with the ones for the matter fields $\phi^m$
\be\lb{pleto2}
E^m(g_{\mu\nu}, \partial_\mu g_{\mu\nu}, \phi^m, \partial_\mu \phi^m, \partial_\mu\partial_\nu \phi^m)=0.
\ee
As (\ref{ee2}) and (\ref{ee3}) determine the curvature functional form as $R_{\mu\nu}=R_{\mu\nu}(g_{\mu\nu}, \partial_\alpha g_{\mu\nu}, \partial_{\alpha}\partial_\beta g_{\mu\nu})$, the equations (\ref{pleto1})-(\ref{pleto2}) are a system of equations for finding $g_{\mu\nu}$ and the matter fields $\phi^m$ as a functions on the space time. In these terms, the solutions of the model can been found. The problem is that the condition of zero torsion tacitly assumed implies, by use of (\ref{tordes}), that 
$$
C_{\mu\nu\alpha}=C_{\alpha\nu\mu},
$$
and therefore $C_{\mu\nu\alpha}$ is completely symmetric under the interchange of any pair of indices. This gives a set of relations that is hardly satisfied by the already determined solution. Note that the number of relations grow with the number or dimensions. Therefore, it is likely that the torsion should be turned on for finding solutions, otherwise this inconsistency will appear.

\section{Discussion}
In the present work, the theories described by a lagrangian of the functional form  $L=L(R_{(\mu\nu)})$ with $R^\mu_{\nu}=g^{\alpha\beta}R^\mu_{\alpha\nu\beta}$ were considered, by finding the equations of motion corresponding to the Palatini formalism.
This complements the studies of reference \cite{olmo}, which considers the choice $\widetilde{R}_{\mu\nu}=R^\rho_{\mu\rho\nu}$. These choices are physically inequivalent and the resulting theories are different. In particular, in the type of theories considered in \cite{olmo}, there is a projective symmetry that allows to dispose the torsion field. In the present type of models instead, it is likely that there is not such a symmetry. It was argued in the text that in the present case the torsion is non trivial and may in fact be healthy for the resulting equations to be well posed. It may be  an interesting task to understand causality issues about the simpler two dimensional scenario described above, the existence and uniqueness of solutions, the role of diffeomorphisms and the existence or absence of superluminal modes.

When this paper was finished, we realized about the existence of the reference \cite{borunda} which  have some partial overlap with the present work. However, this reference emphasize the circumstances in which the torsion results trivial, and furthermore, in which situations the resulting connection is the Levi-Civita one. Some preliminary examples were found in \cite{borunda2}. The present work instead is more focused in characterizing the torsion, at least implicitly. The fact that these theories posses an intrinsic torsion if of interest, specially when fermions are turned on. The way that fermions may couple to the gravitational field in the model presented here may lead to interesting effects to be studied further. Examples can be found in the extensive reviews \cite{shapiro}-\cite{torsion2} and references therein. But it should be mentioned that a contact spin torsion interaction can have phenomenological consequences at high energies such as the ones corresponding the early universe. This may have influence, for instance, in primordial nucleosynthesis. Furthermore, there exist experiments employing polarized media and bodies in order to detect torsion \cite{experimentos}. In addition, axial-vector type interactions between matter field interactions may give forward-backward asymmetry of particle scattering. Further consequences in particle physics are described, for instance, in \cite{dubna}. Another interesting feature is that the presence of torsion may lead to a singularity avoidance \cite{shapiro}. There exists models with effective energy density and pressure due to torsion contributions, for which the torsion plays the role of quintessence \cite{capo}. Further applications in cosmology may be considered following, for instance, references \cite{nikodem1}-\cite{bijan2} and \cite{torre1}-\cite{Vasak}. 

As a final remark we would like to emphasize that the solution for the torsion found in the previous sections for the $1+1$ dimensional case may have interesting applications, even taking into account its low dimensionality. In order to motivate this, recall that the four dimensional chiral anomalies can be understood by studying the Dirac sea in two dimensions \cite{rubakov}. In the same spirit, it may be of interest to consider two dimensional models for fermions as a toy model for domain wall interactions with such particles. In particular, if torsion effects may have some influence on phenomena such as CP violation or baryogenesis. Of course, due to the non linearity of the gravitational field, it is not totally valid to describe a  four dimensional scenario with two translational spatial symmetries with a two dimensional gravitational model. Nevertheless, to do so may serve for gaining intuition about qualitative aspects of these effects. In this context, the models presented here may be of particular interest, and further studies about their solutions may be  worthwhile.

\section*{Acknowledgments}
Both authors are supported by CONICET, Argentina.

  \end{document}